\documentclass[preprint]{article}
\usepackage{psfig,a4wide,amsmath,amsfonts}
\begin{document}

\baselineskip 8mm

\begin{center}  {\Large
 A Quantum Fluctuation Theorem}

\vspace*{1mm}

 {\Large  Jorge Kurchan }

 {\it P.M.M.H. Ecole Sup\'erieure de Physique et Chimie\\
Industrielles, 10 rue Vauquelin 75231 Paris, France
}

\end{center}

\begin{abstract}
We consider a quantum system strongly driven by forces that are
periodic in time.
The  theorem concerns the probability $P(e)$ of observing a given
energy change $e$ after a   number of cycles.
If the system is thermostated by a 
(quantum) thermal bath, $e$ is the
total amount of energy transferred to the bath, while for an isolated system
 $e$ is the increase in energy of the system itself.
Then, we show that
$P(e)/P(-e)=e^{\beta e}$, a parameter-free, model-independent
relation.

\end{abstract}
%\twocolumn 

%\narrowtext

%\baselinestretch{1.4}

In the past few years there has been a renewed interest in the study of
quantum
systems out of equilibrium, to a large extent stimulated by the 
design  of new experimental settings and by the construction  of new devices. 
If a system is well out of equilibrium, as for example when it is
strongly driven by periodic forces, then linear response theory
(understood as linear perturbations around the Gibbs measure) is insufficient.

Even in the context of classical mechanics not many generic results are
available  beyond linear response.  
An interesting   new development consists of a
number of  relations for strongly out of equilibrium systems, 
mainly regarding  the distribution of work and entropy production.
The first of such {\em fluctuation theorems} was discovered  by Evans et.al 
\cite{Evans},   who  understood that the basic ingredient was
time-reversal symmetry.

Two important   further steps, made by 
 Gallavotti and Cohen widened  the scope and 
interest of the subject. On the one hand, it was realized
  \cite{Ga} that the fluctuation  theorems are indeed the 
far from equilibrium generalisations  of the well-known equilibrium theorems 
(fluctuation-dissipation and Onsager reciprocity).
Most intriguingly, a byproduct of their proof  \cite{Gaco} was that,
 just as the validity of the
fluctuation-dissipation relation
is a strong indication of
equilibration,
 the fact that a fluctuation formula holds in a
driven stationary system  strongly hints that the system can be
 considered
`as ergodic as possible' --- with all the implications this entails. 

These results concern deterministic systems. 
If instead a finite system is in contact with a stochastic thermal bath,
then the `ergodicity' questions become trivial, and all the fluctuation
formulae are extremely simple to prove \cite{Ku,Lesp,maes}. 

Particularly relevant for the present work are  the simple and remarkable
Jarzynski and other {\em `work relations'} valid well out of equilibrium 
\cite{Ja,Cr1,related,Yu},
whose close relation to the fluctuation theorems was clarified by
Crooks \cite{Cr2}.  

Except for Ref.\cite{Yu},
the developments described so far are restricted to classical mechanics.
The purpose of this paper is to prove two versions of a fluctuation
theorem  for quantum systems under strong periodic drive, either
isolated or in contact with a thermal bath.
Very recently, these questions have become relevant in the context of
detection of quantum `shot' noise generated by currents flowing
through devices, given the possibility it offers to observe different
quasiparticle charges of the carriers \cite{joseph}.

We shall consider an evolution generated by a  time-dependent
Hamiltonian $H(q,p,t)$ with {\em real} matrix elements, so that 
there is time-reversal symmetry in each infinitesimal time step.
 We shall assume the explicit time-dependence
is periodic (with period $t_o$) and that the cycles are symmetric
\cite{Chi}:
\begin{equation}
H(q,p,t)=H(q,p,t)=H(q,p,-t)
\label{uno}
\end{equation}
Splitting the evolution in infinitesimal steps
$t_1,\;...\;t_n$:
 \begin{equation}
 |\phi(t_f)\rangle = U|\phi(t_i)\rangle=
 U_n U_{n-1} \;\;...\;\; U_2 U_1 |\phi(t_i)\rangle
\label{evol}
\end{equation}
where $n$ is large, $U_r\equiv e^{\frac{i}{\hbar} \tau H_r}$ and 
$\tau=(t_f-t_i)/n \rightarrow 0$, 
the reality condition implies that:
$
U_r^\dag = U_r^*
$
and, together with the symmetry of cycles Eq.  (\ref{uno}) this implies
 that the  evolution over a cycle satisfies time-reversibility:
Consider the case in which the time $t_f-t_i$ consists of an
integer number of cycles, and hence $U_r=U_{n-r+1}$
 (cfr. (\ref{uno})). Then:
\begin{eqnarray}
U^\dag&=&  U_1^\dag U_{2}^\dag \;\;...\;\; U_{n-1}^\dag U_n^\dag=
\left[ U_1 U_{2} \;\;...\;\; U_{n-1} U_n\right]^* \nonumber\\
&=&
\left [ U_n U_{n-1} \;\;...\;\; U_{2} U_1\right]^*=U^*
\label{cosa}
\end{eqnarray}
a formula only valid for $(t_f-t_i)/t_o=integer$, which we shall
assume throughout this paper.

We  denote the eigenvectors of the initial time Hamiltonian
 $|\psi_\alpha\rangle$:
 \begin{equation}
H_1 |\psi_\alpha\rangle = \varepsilon_\alpha  |\psi_\alpha\rangle
\end{equation}
and the corresponding partition function:
\begin{equation}
Z = {\mbox{Tr}} e^{-\beta H_1}
\label{zeq}
\end{equation}

Consider the following protocol:

\begin{itemize}

\item

With (canonical) probability $p_\alpha= e^{-\beta  \varepsilon_\alpha}/Z$ we
choose a wavefunction $|\psi_\alpha\rangle$.

\item

We let it evolve through an integer number of cycles:
 \begin{equation}
|\phi \rangle =  U
 |\psi_\alpha\rangle
 = \sum_\gamma |\psi_\gamma \rangle \langle \psi_\gamma |\phi
 \rangle
\end{equation}

\item

We measure the final  $H_1$ and record the 
energy difference $e$ between initial and
final times.

\end{itemize}

Let us calculate the probability distribution of $e$.
For a given value of $\alpha$, 
this distribution   reads:
\begin{eqnarray}
P_\alpha (e)&=& \sum_\gamma  \delta
[e-(\varepsilon_\gamma-\varepsilon_\alpha)]
\left|\langle \psi_\gamma |\phi\rangle\right|^2 \nonumber \\
&=& \sum_\gamma  \delta
[e-(\varepsilon_\gamma-\varepsilon_\alpha)]
\left|\langle \psi_\gamma | U
 |\psi_\alpha\rangle\right|^2
\end{eqnarray}
The average distribution  $P(e)$  over initial conditions is:
\begin{equation}
P(e)= \frac{1}{Z} \sum_{\gamma,\alpha}
 \delta
[e-(\varepsilon_\gamma-\varepsilon_\alpha)]
\left|\langle \psi_\gamma | U
 |\psi_\alpha\rangle\right|^2
 e^{-\beta 
\varepsilon_\alpha}
\label{ppp}
\end{equation}
Writing the delta function in integral form, we get:
\begin{eqnarray}
P(e)&=& \frac{1}{Z} \int_{-i\infty}^{i \infty} d\lambda 
 e^{-\lambda [e-(\varepsilon_\gamma-\varepsilon_\alpha)]} 
\sum_{\gamma,\alpha}
\left|\langle \psi_\gamma | U
 |\psi_\alpha\rangle\right|^2
 e^{-\beta 
\varepsilon_\alpha} \nonumber \\
&=&  \frac{1}{Z} \int_{-i\infty}^{i \infty}d\lambda  e^{-\lambda e}
 Q(\lambda)
\label{peq}
\end{eqnarray}
where:
\begin{eqnarray}
 Q(\lambda)&=& \sum_{\gamma,\alpha} 
 e^{\lambda (\varepsilon_\gamma-\varepsilon_\alpha)} 
\left|\langle \psi_\gamma | U
 |\psi_\alpha\rangle\right|^2
 e^{-\beta \varepsilon_\alpha} \nonumber \\
&=&  \sum_{\gamma,\alpha}  e^{\lambda \varepsilon_\gamma}
\langle \psi_\alpha | U^\dag   
 |\psi_\gamma\rangle \langle 
\psi_\gamma | U
 |\psi_\alpha\rangle e^{-(\lambda+\beta) (\varepsilon_\alpha)} 
\nonumber \\
&=& {\mbox{Tr}} \left[
U^\dag \; 
 e^{\lambda H_1} U \; e^{-(\lambda+\beta)H_1}
\right]
\label{qeq}
\end{eqnarray}

 The fluctuation theorem can be proved very simply. Firstly, 
let us prove the following KMS-like \cite{Kubo} relation. Let $D$ be
any {\em real symmetric} operator. Then:
\begin{eqnarray}
 Q_D(\lambda) &\equiv& {\mbox{Tr}} \left[ U^\dag e^{\lambda D} U 
e^{-(\lambda+\beta)D} \right]=
 {\mbox{Tr}} \left[
U^\dag
 e^{-(\lambda+\beta)D}
U
 e^{\lambda D} 
\right] \nonumber \\
&=& Q_D(-\lambda^*-\beta)^* = Q_D(-\lambda-\beta)
\label{qeq2}
\end{eqnarray}
where we have used transposition, cyclic permutation and Eq. (\ref{cosa}).
Putting $D=H_1$, this is a form of the fluctuation theorem for 
$Q(\lambda)=Q_D(\lambda)$.

In order to see what the implications of (\ref{qeq2}) are for $P(e)$
we shall need to show that $Q(\lambda)$ is analytic on the stripe
$-\beta \leq  {\mbox{Re}}(\lambda) \leq 0$.
To do this, we use the fact that:
\begin{equation} 
\left| {\mbox{Tr}} AB\right|^2 \leq 
2 \left( {\mbox{Tr}} AA^\dag\right) \left( {\mbox{Tr}} BB^\dag\right)
\end{equation}
Putting $A= 
 U^\dag e^{\lambda H_1}$ and 
$B=U e^{-(\lambda+\beta)H_1}$:
\begin{equation} 
\left| Q(\lambda)\right|^2 \leq
2 \left( {\mbox{Tr}} e^{-2\left\{{\mbox{\small{Re}}}(\lambda)+\beta
\right\} H_1} \right)
 \left( {\mbox{Tr}}  e^{2 {\mbox{\small{Re}}}(\lambda) H_1}\right)
\label{bound}
\end{equation}
Because the partition function (\ref{zeq}) converges for positive
temperatures, neither factor diverges if 
$-\beta <  {\mbox{Re}}(\lambda) < 0$

Inserting
 (\ref{qeq2}) with $D=H_1$ in (\ref{peq}), the  analyticity  result
allows us to shift the integration from ${\mbox{Re}}(\lambda) =0$ to
 ${\mbox{Re}}(\lambda)= -\beta$, and we get:
\begin{equation} 
P(e)=P(-e)e^{\beta e}
\label{ft2}
\end{equation}
This is the fluctuation theorem for the probability of energy
changes  in an isolated system.
Note that we could have obtained this result directly, without
writing the KMS equation, by using the (time-reversal) 
symmetry of $U$ (equation (\ref{cosa})) in (\ref{ppp}).

Taking the expectation value of $e^{-\beta e}$ over an integer number
of cycles, one obtains a 
quantum version of a Jarzynski work formula \cite{Yu}:
\begin{equation} 
{\overline{e^{-\beta e}}}  = \int de P(e) e^{-\beta e} = \int de
P(-e) =1
\label{jar}
\end{equation}
where the overline denotes average over quantum amplitudes and 
initial conditions.
This is a rather surprising model-independent result, which we can
 rewrite as:
 \begin{equation} 
0= -\frac{1}{\beta} \ln {\overline{e^{-\beta e}}}  \leq
 -\frac{1}{\beta} {\overline{ \ln e^{-\beta e}}} =  {\overline{ e}}
\label{jar1} 
\end{equation}
We obtain  the second principle, arising as the familiar inequality
 \begin{equation} 
  annealed \;\; average \leq  quenched\;\;  average
\end{equation}
\vspace{.2cm}
with the initial conditions playing the role of disorder.

\vspace{.3in}

{\em Generalisation and Thermostated systems}

\vspace{.3in}

In the most physical setting, we have a system in contact with a bath.
The bath  can be modeled for example  with an infinite set of harmonic
oscillators coupled to each variable in the system.
Consider for simplicity the case of one system variable:
\begin{equation}
H(x,p_x,y_1,p_1,...,y_M,p_M)=H_{system}+ H_{int}+H_{bath}
\label{bath}
\end{equation}
with:
\begin{equation}
H_{bath}= \sum_i \left[ \frac{p_i^2}{2m} + m \omega_i^2 \frac{y_i^2}{2} \right]
\end{equation}
\begin{equation}
H_{int}= \sum_i C_i y_i x
\end{equation}
and, say:
\begin{equation}
H_{system}= \frac{p_x^2}{2m_x} + V(x,t)
\end{equation}
(The interested reader will find an extensive literature on this
implementation of heat baths in \cite{bath} and references therein.)

We can define the  following bases of eigenvalues:
\begin{eqnarray}
H_{bath}| \chi_\alpha \rangle &=& \varepsilon^y_\alpha
|\chi_\alpha \rangle \nonumber \\
O_{system}| \psi_{\alpha'} \rangle&=& o_{\alpha'} | \psi_{\alpha'} \rangle 
\end{eqnarray}
$O$ is any real  symmetric operator corresponding to an observable
depending  {\em exclusively} on the system variables, with a spectrum bounded 
from below ({\em e.g.}  $H_{system}(t_i)$). 
The wavefunctions   $|\chi_\alpha \rangle$ and  $| \psi_\alpha \rangle$ 
depend only on  bath and system variables, respectively.

We consider an initial condition with no correlations between bath and system
 constructed as follows:

\begin{itemize}

\item

We choose a bath wavefunction 
 $| \chi_\alpha \rangle$ with canonical probability
$\propto e^{-\beta \varepsilon^y_\alpha}$.

\item

We choose a system wavefunction  $| \psi_{\alpha'} \rangle$ with probability
$p_{\alpha'} \propto e^{-\beta o_{\alpha'}}$. 
This distribution is quite general,  given the freedom 
of choice of $O$. It could be for example a canonical distribution at
higher temperature ($O=\frac{\beta'}{\beta} H_1$), as resulting from 
a temperature  quench. 

\item
We start with the initial state $| \chi_\alpha \rangle \otimes
| \psi_{\alpha'} \rangle $ and let it evolve.

\item
We measure
\begin{equation} 
D \equiv H_{bath}+O
\label{momo}
 \end{equation} 
at the beginning and at the end, and record the 
difference.
Note that at this stage the operator $O$ which we are measuring 
is forced by the initial condition we are choosing.

\end{itemize}

It is now easy to prove that the fluctuation formula (\ref{ft2}) holds, with
$e$ measuring the difference in value of  bath energy plus observed value of 
$O$.
To do this one uses (\ref{qeq2}) with $D$ as in (\ref{momo}). 
In order to guarantee the 
analyticity of $Q_D$, one must assure that
 (see argument leading to (\ref{bound}))
\begin{eqnarray}
{\mbox{Tr}}\left[ e^{-2\left\{{\mbox{\small{Re}}}(\lambda)+\beta
\right\} D}\right]
&=&{\mbox{Tr}}\left[ e^{-2\left\{{\mbox{\small{Re}}}(\lambda)+\beta
\right\} H_{bath}}\right]
\nonumber \\
&\times& {\mbox{Tr}}\left[ e^{-2\left\{{\mbox{\small{Re}}}
(\lambda)+\beta
\right\} O}\right] < \infty
\end{eqnarray}
and 
\begin{equation}
 {\mbox{Tr}} \left[  e^{2 {\mbox{\small{Re}}}(\lambda) D}\right]=
{\mbox{Tr}}\left[   e^{2 {\mbox{\small{Re}}}(\lambda) H_{bath}}\right]
{\mbox{Tr}}\left[   e^{2 {\mbox{\small{Re}}}(\lambda) O}\right]< \infty
\end{equation}
 for $-\beta <  {\mbox{Re}}(\lambda)  <  0$.
The traces over the bath variables are bounded, since
they correspond to the bath partition function.
 For the initial probabilities of the
system, 
these conditions require that:
\begin{equation}
{\mbox{Tr}} \left[   e^{-2\beta \mu  O}\right] \propto
\sum_\alpha p_\alpha^{2\mu}
\end{equation}
is finite for $0<\mu<1$, a condition we assume.

We are now in a position of discussing a general system in contact
with a an infinite 
heat bath after many cycles have elapsed ($t_f-t_i \rightarrow
\infty$).  If we have that, under these circumstances:

 {\em i)}  The expected value of $O$ --- a property of the system --- 
 stays finite as $t_f$ grows.

 {\em ii)}  The bath  receives by virtue of the time-dependent forces
an energy proportional  to the number of
cycles \cite{foot}: $e = (t_f-t_i) e_o + e_1$, with $e_1$ finite and
and $e_o$ {\em independent of the initial configuration}.

Then, the derivation above of the fluctuation theorem
carries over for long times to {\em any} initial distribution
and
 to leading order in $t_f-t_i$,
 equation (\ref{ft2})  will be a statement about the probability
distribution of the  energy $e_o$ 
the bath received, i.e. its entropy increase \cite{ffot}.
Measuring the fluctuation formula in a concrete situation becomes then
a test for a property of the `stationary' (i.e. periodical) asymptotic
quantum state.

\vspace{.5cm}

\vspace{.5cm}

{\bf Acknowledgements}

\vspace{.2cm}

I wish to thank L.F. Cugliandolo, D. Grempel 
and P. Leboeuf for useful suggestions.  
I am indebted to Y. Imry for calling my attention to the connections with
works on quantum devices.


\begin{thebibliography}{99}
\bibitem{foot} For this to happen there should be no `localisation' 
in energy, as can happen in 
an isolated quantum  `kicked' system\cite{Chi}.
\end{thebibliography}
\end{document}